\documentclass[a4paper]{PoS}

\newcommand{\nuebar}{$\overline{\nu}_{e}$}

\title{Estimation of reactor neutrino fluxes}

\ShortTitle{Estimation of reactor neutrino fluxes}

\author{\speaker{Daniel A. Dwyer}\\%\thanks{A footnote may follow.}\\
        Lawrence Berkeley National Laboratory\\
        E-mail: \email{dadwyer@lbl.gov}}

\abstract{

  Reactor antineutrinos have been indispensable for our understanding
  of neutrino mass and mixing.
  At the same time, discrepancies between the observed and predicted
  reactor \nuebar{} rate and energy spectra have grown as the
  precision of these measurements has improved.
  Measurements of the electrons emitted following fission result in
  the most precise predictions for the corresponding \nuebar{} flux,
  and our understanding of the potential systematic differences between
  the fission $e^-$ and \nuebar{} fluxes has improved.
  Measurements of individual fission daughter isotopes and their
  decays are fraught with uncertainties, yet still provide insight
  into these discrepancies.
  Detailed comparisons of \nuebar{} measurements among reactors are
  also shedding new light on this topic.
}

\FullConference{Neutrino Oscillation Workshop\\
		4 - 11 September, 2016\\
		Otranto (Lecce, Italy)}

\begin{document}

\section{Introduction}

Antineutrinos emitted by nuclear fission reactors have served as a
powerful tool for the study of these weakly-interacting particles.
The intense flux from reactors, roughly 10$^{20}$ \nuebar{} per second
per GW$_{\mathrm{th}}$ of reactor power, was used for the first
detection of these elusive particles~\cite{Reines:1960pr}.
Measurements of reactor antineutrinos have also revealed the distinct
signature of the oscillation of neutrino flavor~\cite{Araki:2004mb,
  An:2015rpe}.
On the other hand, precise models of reactor \nuebar{} emission do not
agree with these measurements.
The predicted rate is 6\% higher than that observed, a feature that is
commonly referred to as the {\em reactor antineutrino anomaly} and has
been considered possible evidence for sterile
neutrinos~\cite{Mention:2011rk}.
More recently, precise measurements of \nuebar{} energy spectra have
also shown a $\sim$10\% excess relative to prediction in the region of
5 to 7~MeV~\cite{An:2016srz, Seo:2016uom, Cabrera:Nu2016}.
In these proceedings I will examine the details behind these
discrepancies, and discuss the substantial recent developments in this
field.

%\section{Reactor \nuebar{} production}

The process of reactor \nuebar{} production is well understood.
Fission of actinides, in particular $^{235}$U, $^{238}$U, $^{239}$Pu,
and $^{241}$Pu, produce unstable neutron-rich fission fragments.
These fission daughter isotopes undergo successive $\beta$-decays
until reaching stability, with an average of 6 decays per initial
fission.
The total \nuebar{} emission from a reactor, $S(E_{\overline{\nu}})$,
is the sum of the \nuebar{}'s emitted by these decays,
\begin{equation}\label{eq:totalSpectrum}
  S(E_{\overline{\nu}}) = \sum\limits_{i=0}^{n}R_i\sum\limits_{j=0}^{m}f_{ij}S_{ij}(E_{\overline{\nu}}),
\end{equation}
where $R_i$ is the rate of decays of the $i$'th fission daughter
isotope, $f_{ij}$ is the relative probability for the $j$'th decay
mode of this daughter (also referred to as the branching fraction),
and $S_{ij}(E_{\overline{\nu}})$ is the \nuebar{} energy spectrum for
the $j$'th decay mode.
There are more than 1300 known fission daughter isotopes, and combined
they include more than 10,000 unique decay modes.

\section{Current measurements}

Three sets of measurements are particularly relevant to the assessment
of reactor \nuebar{} production:
\begin{enumerate}
\item direct measurements of reactor \nuebar{} emission,
\item measurements of electron emission following fission, and
\item measurements of the fission yields and decay modes of fission daughters.
\end{enumerate}
Direct measurements commonly involve \nuebar{} detection via inverse
beta decay (IBD) in large organic scintillator detectors.
Calorimetry of the positrons produced by IBD allow accurate estimation
of the rate and energy spectra of those \nuebar{} with energies above
the interaction threshold of 1.8~MeV\@.
Subsequent detection of the neutrons produced by IBD allows for
effective background rejection.
The most recent generation of direct measurements have observed more
than 1~million \nuebar{} interactions, and obtained percent-level
uncertainties in both the rate and energy spectra~\cite{An:2016srz,
  Seo:2016uom, Cabrera:Nu2016}.
This precision has been putting pressure on the field to obtain more
accurate predictions of the reactor \nuebar{} flux.

The most precise predictions have been based on corresponding
measurements of the rate and energy spectra of electrons emitted
following fission.
Due to the kinematic symmetry of the electrons and \nuebar{} produced
in $\beta$-decay, their rate and energy spectra are highly correlated.
The fission electron spectra were measured at the 2\%-level in a
series of experiments at the ILL research reactor in Grenoble in the
1980's~\cite{VonFeilitzsch:1982jw, Schreckenbach:1985ep, Hahn:1989zr,
  Haag:2014kia}.
In these measurements, foils of actinides ($^{235}$U, $^{239}$Pu,
$^{241}$U) were exposed to the neutron flux in the ILL reactor and the
emitted electrons were measured.
By measuring the cumulative electron spectra due to all the fission
daughters and their decay modes, one avoids the need to know the
detailed aspects of each daughter.
Modeling the electron spectrum as the sum of a large number of
$\beta$-decays, the corresponding \nuebar{} spectrum can be
calculated~\cite{Vogel:2007du}.
Nuclear corrections to $\beta$-decay do introduce slight asymmetries
between the electron and \nuebar{} spectra, as summarized
in~\cite{Huber:2011wv}.
A hybrid approach that uses data on fission daughters to inform these
nuclear corrections gives a similar result~\cite{Mueller:2011nm}.
Overall, this {\em $\beta$-conversion} approach provides a prediction
for the reactor \nuebar{} rate and energy spectra with uncertainties
at the 3\%-level, and has served as the de-facto standard for the past
thirty years.

Unfortunately, the direct \nuebar{} measurements and
$\beta$-conversion predictions disagree on both the rate and energy
spectra, as discussed in the introduction.
The origin of these discrepancies are unclear, although potential
explanations have been explored~\cite{Hayes:2015yka}.
Antineutrinos from the decay of neutron-activated reactor materials,
spectral distortions from forbidden decays, and non-thermal fission of
$^{238}$U do not seem to be large enough to explain the differences.
The energy spectra of the neutrons producing fission in the ILL
electron measurements differ slightly from that in the commercial
reactors used in the direct \nuebar{} measurements.
This could result in a slightly different distribution of fission
daughters, which is difficult to rule out as a source for the
discrepancies between the electron and \nuebar{} data.
Another option could be an unknown systematic in the ILL electron
measurements, although this is difficult to confirm given that these
are the only set of electron measurements to date.

What guidance can the past century of measurements of nuclear fission
and decay provide?
These measurements, which are collected in nuclear databases such as
ENDF, JEFF, and JENDL, can be used to calculate the \nuebar{} flux
according to Eq.~\ref{eq:totalSpectrum}.
Examples of such calculations can be found in~\cite{Vogel:1980bk,
  Fallot:2012jv, Sonzogni:2015aoa, Dwyer:2014eka}.
Given the large uncertainties of such calculations, one might conclude
that these databases can provide little guidance.
For example, 70\% of the known fission daughters lack decay mode data
(although these tend to be those daughters which are rarely produced,
and hence only amount to $\sim$6\% of the total fission yield).
The fission yield data provided by the various databases are
inconsistent with each other, and gross errors have been
identified~\cite{Sonzogni:2016yac}.
Decay data are generally only known for the most prominent decay
modes, and are susceptible to systematic biases from measurement
techniques (e.g.\ the Pandemonium effect).%~\cite{Hardy:1977bd}.

Despite these obstacles, the shape of reactor \nuebar{} energy
spectrum calculated from the ENDF database is unexpectedly similar to
the direct \nuebar{} measurements~\cite{Dwyer:2014eka}.
This may not be wholly surprising, since the spectral shape seems to
be dominated by a small number of prominent fission daughters and
decay modes which are well-measured.
Many of the uncertainties impact daughters and modes which each
contribute at most 1\% of the overall \nuebar{} flux, and hence have
little influence on the spectral shape.
Consequently, there is potential to improve the calculation of the
spectral shape through a targeted program of measurement of the most
prominent fission daughters.
Unfortunately, the rate calculation will likely continue to suffer
from large uncertainties due to the cumulative effect of the many rare
but poorly known fission daughters.

\section{Looking forward}

A targeted program of measurements of the decay modes of prominent
fission daughters is being pursued, and has begun to yield results.
In particular, measurements of $^{92}$Rb and $^{142}$Cs using total
absorption spectroscopy have already reduced the largest uncertainties
in the calculation of the 5 to 7~MeV discrepant region of the
\nuebar{} spectrum~\cite{Zakari-Issoufou:2015vvp, Rasco:2016leq}.
Comparison of the ENDF and JEFF databases suggest another important
step will be improved measurements of the fission yields of the most
prominent daughters, of which $^{96}$Y is the most
critical~\cite{Hayes:2015yka}.
To directly address the tension between the electron and \nuebar{}
measurements, a repeat of the ILL electron measurements is being
considered at LANL.

Recent work comparing the direct \nuebar{} measurements between
different nuclear reactors has also been fruitful.
A global analysis of \nuebar{} rate measurements has shown that the
rate discrepancy cannot be attributed solely to the minor fission
parents such as $^{239}$Pu or $^{238}$U, and instead shows that
$^{235}$U electron and \nuebar{} data are in
tension~\cite{Giunti:2016elf}.
A double ratio of the Daya Bay and NEOS observed over expected
\nuebar{} spectra also suggests tension between the $^{235}$U electron
and \nuebar{} energy spectra~\cite{Huber:2016xis}.
Data from the upcoming generation of short-baseline direct \nuebar{}
measurements, such as PROSPECT~\cite{Ashenfelter:2015uxt}, should
continue to elucidate.
The impressive precision of recent \nuebar{} measurements also
suggests interesting potential for reactor characterization and
non-proliferation.
In general, our understanding of reactor \nuebar{} emission is
advancing rapidly and I expect substantial improvements over the
coming years.

% Acknowledgements
I would like to thank the organizers of the 2016 Neutrino Oscillation
Workshop for the invitation to come speak on this topic.
I owe Patrick Huber, Bryce Littlejohn, and Patrick Tsang for
thoughtful discussions on these topics.
This work was supported under DOE OHEP DE-AC02-05CH11231.

%\begin{thebibliography}{99}
%\bibliographystyle{unsrt}
%\bibliography{neutrino}

%\end{thebibliography}

\end{document}